\documentclass[aps,prl,twocolumn,superscriptaddress]{revtex4}

\usepackage{graphicx}
\usepackage{amsmath}
\usepackage{amssymb}

\begin{document}
\title{Origin of the large polarization in multiferroic YMnO$_3$ thin films 
revealed by soft and hard x-ray diffraction}

\author{H.~Wadati}
\email{wadati@ap.t.u-tokyo.ac.jp}
\homepage{http://www.geocities.jp/qxbqd097/index2.htm}
\affiliation{Department of Applied Physics and Quantum-Phase Electronics 
Center (QPEC), University of Tokyo, Hongo, Tokyo 113-8656, Japan} 

\author{J. Okamoto}
\affiliation{Condensed Matter Research Center and Photon Factory,
Institute of Materials Structure Science, High Energy Accelerator
Research Organization, Tsukuba 305-0801, Japan}

\author{M. Garganourakis}
\affiliation{Swiss Light Source, Paul Scherrer Institut, 
5232 Villigen PSI, Switzerland} 

\author{V. Scagnoli}
\affiliation{Swiss Light Source, Paul Scherrer Institut, 
5232 Villigen PSI, Switzerland} 

\author{U. Staub}
\affiliation{Swiss Light Source, Paul Scherrer Institut, 
5232 Villigen PSI, Switzerland} 

\author{Y. Yamasaki}
\affiliation{Condensed Matter Research Center and Photon Factory,
Institute of Materials Structure Science, High Energy Accelerator
Research Organization, Tsukuba 305-0801, Japan}

\author{H.~Nakao}
\affiliation{Condensed Matter Research Center and Photon Factory,
Institute of Materials Structure Science, High Energy Accelerator
Research Organization, Tsukuba 305-0801, Japan}

\author{Y. Murakami}
\affiliation{Condensed Matter Research Center and Photon Factory,
Institute of Materials Structure Science, High Energy Accelerator
Research Organization, Tsukuba 305-0801, Japan}

\author{M. Mochizuki}
\affiliation{Department of Applied Physics and Quantum-Phase Electronics 
Center (QPEC), University of Tokyo, Hongo, Tokyo 113-8656, Japan}

\author{M. Nakamura}
\affiliation{Cross-Correlated Materials Research Group (CMRG), 
RIKEN Advanced Science Institute, Wako 351-0198, Japan}

\author{M. Kawasaki}
\affiliation{Department of Applied Physics and Quantum-Phase Electronics 
Center (QPEC), University of Tokyo, Hongo, Tokyo 113-8656, Japan} 
\affiliation{Cross-Correlated Materials Research Group (CMRG), 
RIKEN Advanced Science Institute, Wako 351-0198, Japan}

\author{Y. Tokura}
\affiliation{Department of Applied Physics and Quantum-Phase Electronics 
Center (QPEC), University of Tokyo, Hongo, Tokyo 113-8656, Japan} 
\affiliation{Cross-Correlated Materials Research Group (CMRG), 
RIKEN Advanced Science Institute, Wako 351-0198, Japan}


\date{\today}
\begin{abstract}
%Thin films of multiferroic perovskite $R$MnO$_3$ ($R$: rare-earth) 
%are important for device applications. 
We investigated the 
magnetic structure of an orthorhombic YMnO$_3$ thin film 
by resonant soft x-ray and hard x-ray diffraction. 
We observed a temperature-dependent incommensurate magnetic reflection 
below 45 K and a commensurate lattice-distortion reflection below 35 K. 
These results demonstrate that the ground state is 
composed of coexisting E-type and cycloidal states. 
Their different ordering temperatures clarify 
the origin of the large polarization 
to be caused by the E-type 
antiferromagnetic states 
in the orthorhombic YMnO$_3$ thin film. 
\end{abstract}
\maketitle
%\section{Introduction}
Recently, there has been a lot of interest 
in multiferroic materials, which display 
both ferroelectric and magnetic orders 
with giant magnetoelectric coupling 
\cite{tokuramulti, cheong, Sekitokura}. 
It is of particular importance to control 
magnetization (electric polarization) 
by electric (magnetic) field for novel 
device applications. 
Orthorhombic ($o$-) $R$MnO$_3$ ($R$: rare-earth) 
with perovskite structure are prototype 
multiferroic materials. For example, 
in TbMnO$_3$, ferroelectricity occurs below 28 K, 
concomitant with the onset of cycloidal spin 
ordering \cite{kimuratmo,kimuratmoprb,Kenzelmann}. 
The ferroelectricity in the cycloidal states 
is realized by the shifts of the oxygen ions 
through the inverse Dzyaloshinskii-Moriya 
interaction \cite{katsura,Mostovoy}. 
This is in contrast to 
E-type antiferromagnetic structures 
($\uparrow\uparrow\downarrow\downarrow$ 
type), where ferroelectricity is caused 
by symmetric exchange striction \cite{sergienko}. 
E-type magnetic structures occur in $o$-$R$MnO$_3$ 
with smaller $R$ ions. 
It is predicted that the E-type structure leads 
to a larger polarization, which has been experimentally 
confirmed in the $o$-$R$MnO$_3$ systems 
\cite{pomja,ishiwatamulti}. 

The fabrication of the $o$-$R$MnO$_3$ thin films 
has been especially important for device 
application of the multiferroic materials. 
Moreover, bulk $o$-$R$MnO$_3$ samples 
with smaller $R$ ions 
($R=$ Y, Ho - Lu) can only be synthesized 
under high oxygen pressure \cite{ishiwatamulti}, 
which strongly limits studies on the most 
interesting materials 
due to the absence of significantly large 
high-quality single crystals. 
Recently, Nakamura {\it et al.} reported 
the fabrication of $o$-YMnO$_3$ thin films 
onto the YAlO$_3$ (010) substrate \cite{YMOnakamura}. 
Their thin film showed a ferroelectric transition 
at 40 K with a large saturation polarization of 
0.8 $\mu$C/cm$^2$. The ferroelectric polarization 
could be controlled by magnetic fields, demonstrating 
magnetoelectric behaviors. 
%As for YMnO$_3$, 
%the hexagonal phase is stable 
%at ambient pressure, and bulk $o$-YMnO$_3$ 
%can be synthesized only under high oxygen 
%pressure \cite{ishiwatamulti}, so 
%it has an additional significance to obtain 
%a single crystal thin film of YMnO$_3$. 
 
Therefore it is interesting and important to clarify 
the exact magnetic structure of YMnO$_3$ 
thin films. In this study we use the 
technique of resonant soft x-ray diffraction  
at Mn $2p\rightarrow3d$ edges to obtain 
the information of magnetic ordering in YMnO$_3$ 
thin films. Resonant soft x-ray diffraction 
has recently been 
used to study the magnetic ordering 
in multiferroic TbMnO$_3$ and 
Eu$_{3/4}$Y$_{1/4}$MnO$_3$ 
\cite{forest,tmowilkins,RSXS2011} using 
single crystals for the larger $R$-ion 
orthorhombic $R$MnO$_3$ series. 
This technique is especially suitable for 
studying magnetism in thin films 
(as demonstrated on $R$NiO$_3$ \cite{valerio}) 
because even small sample volume of thin films 
can be used due to the large resonant enhancement 
of magnetic scattering at the 
%because a small volume of thin films 
%is not a problem due to a 
%resonant enhancement at the 
transition-metal $2p\rightarrow3d$ edges. 
We detect (0 $q_b$ 0) 
($q_b\sim0.5$) magnetic peak, and 
observed temperature-dependent 
incommensurabilities. 
%From hard 
%x-ray diffraction, we also observed 
%commensurate lattice-distortion peaks. 
From hard x-ray diffraction we found a commensurate 
superlattice reflection (0 1 0) that reflects the lattice distortion 
caused by the E-type magnetic structure. 
These results reveal that 
the ground state of the YMnO$_3$ 
can be described by the coexistence of 
E-type and cycloidal states, 
%is the coexisting phase of 
%E-type and cycloidal states, 
while the E-type state is a dominant 
source for the large electric polarization of 
0.8 $\mu$C/cm$^2$ by the symmetric 
exchange striction. 
%\cite{munoz}

%\section{Experiment}
The thin film (40 nm) of YMnO$_3$ was grown 
on a YAlO$_3$ (010) substrate by pulsed-laser 
deposition. The details of the sample fabrication were 
described elsewhere \cite{YMOnakamura}. 
Resonant soft x-ray diffraction experiments 
were performed on the RESOXS endstation \cite{SLS} 
at the surfaces/interfaces microscopy (SIM) 
beamline of the Swiss Light Source of 
the Paul Scherrer Institut, Switzerland. 
For the azimuthal scans (rotation 
around the Bragg scattering wave vector), 
the sample transfer line was used to rotate the sample holder. 
With pins attached in a threefold symmetry 
on the sample holder, an accuracy of approximately 5 deg 
was obtained. A continuous helium-flow cryostat allows 
measurements between 10 and 300 K. Hard x-ray diffraction 
experiments were performed on beamlines 3A and 4C 
at the Photon Factory, KEK, Japan. The photon energy of 
the incident x-ray was 12 keV. 

%\section{Results and Discussion}
Figure 1 shows the temperature dependence of the (0 $q_b$ 0) 
($q_b\sim0.5$) peak with $\pi$ (a) and $\sigma$ (b) incident x-ray polarizations.
The experimental geometry is shown in Fig.~1 (c), together 
with the definition of the azimuthal angle $\varphi$. 
Here the diffraction data were taken with 
$\varphi=0^{\circ}$ at $h\nu=643.1$ eV 
(Mn $2p_{3/2}\rightarrow$ $3d$ absorption edge). 
We measured in both cooling and heating cycles, and 
observed no hysteresis behavior. 
This peak, which is indicated by vertical bars, 
appears at 45 K, which coincides with the antiferromagnetic 
transition temperature $T_N$ determined 
from magnetization measurements \cite{YMOnakamura}. 
Weaker peaks are observed on both sides of the reflection. 
These are antiferromagnetic Kiessig fringes, and describe 
the limited thickness of the magnetic contrast of the film. 
There is almost no difference 
between $\pi$ (a) and $\sigma$ (b) polarizations. 
The intensity of the peaks increases 
monotonically with cooling. 
The peak position deviates from the commensurate $q_b=1/2$ position 
for all temperatures. The peak position shifts to higher angle 
for decreasing temperatures; 
the temperature variation of 
the corresponding wave vector 
and intensity is summarized in Fig.~2. 
The intensity increases monotonically and smoothly 
with decreasing temperatures from $T_N=45$ K. 
The peak position, e.g. 
$q_b=0.457$ at 44 K and $0.491$ at 11 K, 
is temperature-dependent and always incommensurate ($\ne 1/2$) 
in the temperature range of 11 - 44 K. 
In TbMnO$_3$ the peak position is also 
incommensurate, but lock to 
the value of $q_b=0.285$ at the ferroelectric 
transition temperature $T_C=28$ K \cite{forest,tmowilkins}. 
Such a behavior is not observed in 
this YMnO$_3$ film; there is no locking of the peak position 
at $T_C=40$ K, which was determined from electric 
polarization measurements \cite{YMOnakamura}. 

\begin{figure}
\begin{center}
\includegraphics[width=9cm]{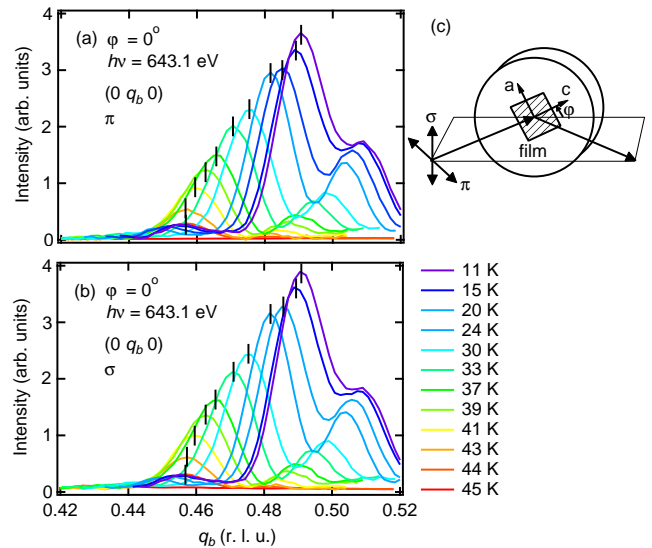}
\caption{(Color online): 
Temperature dependence of the (0 $q_b$ 0) 
($q_b\sim0.5$) peak in $\pi$ (a) and $\sigma$ (b) incident 
x-ray polarizations. 
Panel (c) shows the experimental geometry with the definition 
of the azimuthal angle $\varphi$. In panels (a) and (b), 
the data were taken with $\varphi=0^{\circ}$ at 
$h\nu=643.1$ eV 
(Mn $2p_{3/2}\rightarrow$ $3d$ absorption edge). }
\label{fig1}
\end{center}
\end{figure}

\begin{figure}
\begin{center}
\includegraphics[width=9cm]{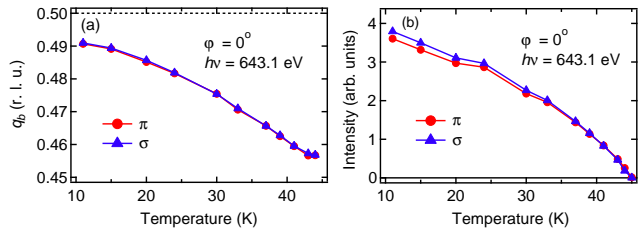}
\caption{ (Color online): 
Temperature dependence of the (0 $q_b$ 0) ($q_b\sim0.5$) 
peak position (a) and intensity (b). The experimental 
geometry and the photon energy are the same as Fig.~1. 
In panel (a), the commensurate position of $q_b=1/2$ 
is shown as a dotted line.}
\label{fig2}
\end{center}
\end{figure}

Figure 3 shows the intensity of the 
(0 $q_b$ 0) ($q_b\sim0.5$) peak 
as a function of photon energies at the 
Mn $2p\rightarrow$ $3d$ absorption edge 
at 44 K (a) and 11 K (b). There is no 
polarization dependence at this scattering geometry 
of $\varphi=0^{\circ}$ at both temperatures. 
In addition, the spectral shape is identical 
at these two temperatures and very similar to the one 
observed for TbMnO$_3$ and 
Eu$_{3/4}$Y$_{1/4}$MnO$_3$ 
\cite{forest,tmowilkins,RSXS2011}. 
This shows that the line shape of the spectrum 
does not depend on the values of $q_b$ but 
is rather common in multiferroic 
$o$-$R$MnO$_3$. 
%because these three materials 
%(YMnO$_3$ thin films, TbMnO$_3$ and 
%Eu$_{3/4}$Y$_{1/4}$MnO$_3$) 
%have different values of $q_b$.
 
\begin{figure}
\begin{center}
\includegraphics[width=9cm]{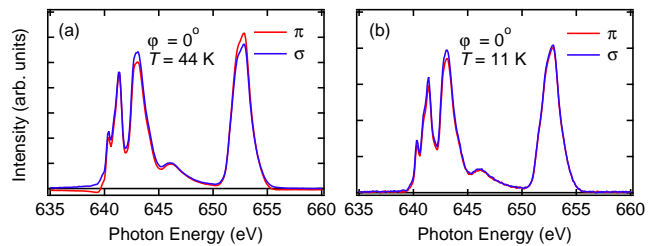}
\caption{(Color online): 
Intensity of the (0 $q_b$ 0) ($q_b\sim0.5$) 
peak as a function of photon energies at the 
Mn $2p\rightarrow$ $3d$ absorption edge 
at 44 K (a) and 11 K (b). }
\label{fig3}
\end{center}
\end{figure}

To gain more information on the spin structure, 
it is important to study the magnetic reflection 
with linear polarized incident radiation 
for different azimuthal angles. 
The $\varphi$ (azimuthal angle) dependence of 
the intensity of the magnetic 
(0 $q_b$ 0) reflection is shown in Fig.~4. 
For $\varphi=0^{\circ}$, 
the intensities are identical for 
$\pi$ and $\sigma$ polarizations within 
experimental uncertainty. 
When $\varphi$ increases from $0^{\circ}$ to 
$90^{\circ}$, the intensity 
increases with incident $\pi$ polarization and 
decreases with incident $\sigma$ polarization. 
The azimuthal-angle dependence allows us 
to gain information on the directions of the Mn spins. 
In the electric-dipole transition, the magnetic 
contribution to the structure factor is given as 
$$
f^{res}_{mag}\propto
(\hat{\epsilon}^{\prime}\times\hat{\epsilon})
\cdot\hat{z}, 
$$
where $\hat{\epsilon}$ and $\hat{\epsilon}^{\prime}$ 
are unit vectors of the incident and scattered polarization, 
respectively, and $\hat{z}$ is a unit vector 
in the direction of the magnetic moment of the ion 
\cite{Hannon,jphill}. We use 
the notations in Fig.~1 in Ref.~\cite{jphill} 
which lead to the following expression, 
$$
(\hat{\epsilon}^{\prime}\times\hat{\epsilon})
\cdot\hat{z}= 
\begin{pmatrix}
0 & z_1\cos\theta_B+z_3\sin\theta_B \\
z_3\sin\theta_B-z_1\cos\theta_B & -z_2\sin 2\theta_B
\end{pmatrix}
$$
Here $\theta_B$ is the Bragg angle for the 
(0 $q_b$ 0) reflection. 
When the magnetic Fourier components contribute 
only along the $c$ axis, 
$z_1=\cos\varphi$, $z_2=\sin\varphi$, and 
$z_3=0$. 
Then the intensity for $\pi$ and $\sigma$ incident 
polarizations are given with $\theta_B\sim51.5^{\circ}$ at 30 K. 
\begin{eqnarray*}
I(\pi) & = & |I(\pi\rightarrow\sigma^{\prime})|^2 + 
|I(\pi\rightarrow\pi^{\prime})|^2\\
 & = & |\cos\varphi\cos\theta|^2 + 
|\sin\varphi\sin 2\theta|^2\\
 & \sim & 0.95-0.56\cos^2\varphi \\
I(\sigma) & = & |I(\sigma\rightarrow\pi^{\prime})|^2 \\
 & = & |\cos\varphi\cos\theta|^2 \\
 & \sim & 0.39\cos^2\varphi
\end{eqnarray*}
The values of these equations are shown as solid lines 
in Fig.~4, and are in good agreement 
with our experimental observations. 
%This indicates that 
%the (0 $q_b$ 0) reflection originates 
%from the spin components along the $c$ axis. 
This reflects an ab cycloid with a spin canting 
along the $c$-axis as shown in Fig.~\ref{fig6}, and indicates 
that the experiment is only 
sensitive to its magnetic sinusoidal $c$ axis component. 

\begin{figure}
\begin{center}
\includegraphics[width=7cm]{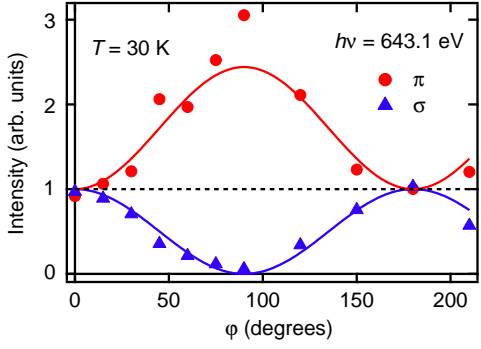}
\caption{(Color online): 
Azimuthal angle dependence of the 
(0 $q_b$ 0) ($q_b\sim0.5$) intensity. 
The solid lines are from the model with 
spins parallel to the $c$ axis. }
\label{fig4}
\end{center}
\end{figure}

\begin{figure}
\begin{center}
\includegraphics[width=8cm]{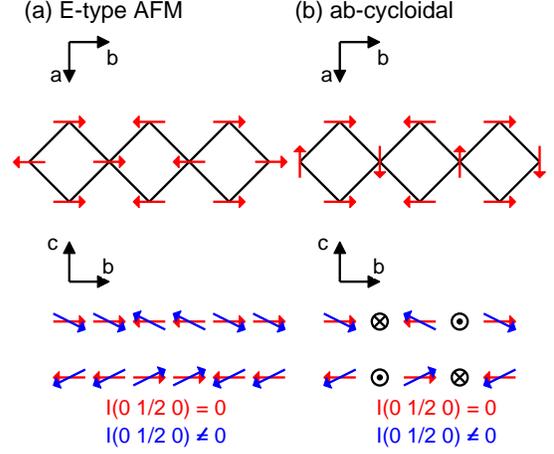}
\caption{(Color online): 
Spin structures in the E-type (a) and the $ab$-cycloidal (b) states. 
Spin canting along the $c$ axis makes the 
magnetic (0 $q_b$ 0) peak have some intensity.}
\label{fig6}
\end{center}
\end{figure}

\begin{figure}
\begin{center}
\includegraphics[width=9cm]{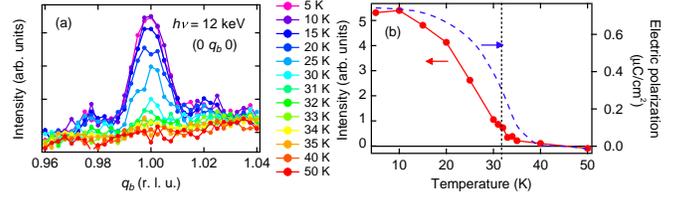}
\caption{(Color online): 
Temperature dependence of the (0 1 0) peak taken 
at $h\nu=12$ keV. In panel (b), peak intensities are 
plotted as a function of temperature together with 
the electric polarization (broken lines) 
taken from Ref.~[12]. The temperature of 35 K is 
also indicated as the onset of the (0 1 0) peak 
and the step onset of the spontaneous electric polarization.}
\label{fig5}
\end{center}
\end{figure}

In order to investigate the lattice distortions associated 
with magnetic order and electric polarization, 
we additionally performed hard x-ray diffraction 
measurements of the YMnO$_3$ thin film. 
The commensurate (0 1 0) reflection appears 
below 35 K as shown in Fig.~\ref{fig5}. 
This reflection is a structurally forbidden 
in the chemical high-temperature structure (Pbnm) 
and caused by the lattice distortion 
accompanying ferroelectricity. 
Interestingly, no incommensurability of this reflection 
is observed by hard x-ray diffraction, 
in clear contrast to the observed magnetic reflection. Moreover, 
this reflection does appear below 35 K, at lower temperatures 
than the onset of the magnetic reflection, 
in accord with the step onset of the spontaneous electric 
polarization [12], as can be seen from the 
temperature-dependent integrated intensity 
shown in Fig.~\ref{fig5} (b). 

We can obtain a full picture of the magnetic states 
of the epitaxial YMnO$_3$ thin film by 
combining the above results with 
the macroscopic measurements of magnetization 
and electric polarization \cite{YMOnakamura}. 
From the macroscopic measurements, 
three transitions were observed: 
antiferromagnetic transition at $T_N=45$ K, 
ferroelectric transition at $T_C=40$ K, 
and an increase of electric polarization 
at 35 K. 
The incommensurate magnetic peak 
was observed at all temperature below 45 K. 
It reflects spin moments solely along the $c$ axis as indicated 
by its x-ray polarization and azimuthal dependence. 
This supports the scenario that in the temperature range of 40 - 45 K 
a sinusoidal state with a spin canting along the $c$ axis is realized. 
Note that the in-plane magnetic moment components cancel 
for this magnetic wave vector in the structure factor. 
This state is also consistent with the absence of observed 
electric polarization in this temperature regime (see Fig.~6 (b)). 
By cooling through 40 K, the sinusoidal magnetic phase transforms 
into a cycloidal magnetic structure with significant magnetic moment 
contributions along the $c$ axis. 
%and the direction of the observed spin 
%is along the $c$ axis. 
%This means that in the temperature range of 
%40 - 45 K, a sinusoidal state with a spin 
%canting along the $c$ axis is 
%realized because such a magnetic ordering 
%has no electric polarization. 
%In the range of 35 - 40 K, 
%ferroelectricity is induced by a cycloidal 
%magnetic structure with a significant 
%magnetic moment contribution along the $c$ axis. 
Below 35 K, we can observe both the incommensurate 
magnetic reflection and the commensurate 
lattice-distortion reflection. This state can be therefore 
explained by the coexistence 
of the cycloidal and the E-type states as theoretically 
predicted in Ref.~\cite{mochi1}. In this coexistence region, 
magnetic reflection is incommensurate as shown 
in Ref.~\cite{mochi1} and lattice peaks are 
commensurate because the E-type phase has a much 
larger lattice distortion than the cycloidal phase. 
The existence of the E-type phase 
causes the large electric polarization of 
0.8 $\mu$C/cm$^2$ due to the symmetric exchange 
striction \cite{YMOnakamura}. 
In other words,  the weak polarization emerging at 40 K 
from the cycloidal magnetic structure causes also weak 
lattice distortion, which is too weak to be observed in our experiment. 
On the other hand, the large induced electric polarization below 35 K 
caused by the E-type structure induces a significant 
lattice distortion, 
as observed by the x-ray diffraction experiments 
on a YMnO$_3$ single crystal \cite{okuYMO}. 
However, spin canting in its magnetic structure 
is so small that 
no additional magnetic contribution is observed in our experiment. 
It is difficult to distinguish between 
the occurrence of $ab$- and $bc$-cycloids 
based on our experimental data. 
However, electric polarization is parallel 
to the $a$ axis \cite{YMOnakamura}, 
which clearly indicates the $ab$-cycloid. 
%The $ab$-cycloid is also consistent with a spin canting 
%along the $c$ axis, where it would 
%make a $bc$-cycloid anisotropic.  
The $ab$-cycloids can easily adopt a spin canting along the $c$ axis, 
whereas $bc$-cycloids would get anisotropically distorted. 

%\section{conclusion}
In summary, we investigated the magnetic structures of 
the YMnO$_3$ thin film by 
resonant magnetic soft x-ray and 
hard x-ray diffraction. 
We observed temperature-dependent 
incommensurate magnetic peaks below 45 K 
and commensurate lattice-distortion peaks 
below 35 K, indicating that 
E-type and cycloidal states coexist below 35 K. 
%Therefore, the E-type phase occurring below 35K is producing 
%a large electric polarization 
%in the YMnO$_3$ thin film. 
This shows that the occurrence of the large electric polarization 
below 35 K is directly related to E-type 
magnetic ordering component in the epitaxial YMnO$_3$ films. 

%\section*{Acknowledgments}
Informative discussions with S.~Ishiwata 
are greatly acknowledged. 
The authors thank the experimental support of the X11MA beamline staff. 
Financial support of the Swiss National Science Foundation and its NCCR 
MaNEP is gratefully acknowledged. This work is also 
supported by the Japan Society for the Promotion 
of Science (JSPS) through its Funding Program for 
World-Leading Innovative R{\&}D on Science and Technology 
(FIRST Program). Hard x-ray diffraction measurements 
were performed under the approval of the Photon Factory Program 
Advisory Committee (Proposals Nos. 2009S2-008 and 2010G678) 
at the Institute of Material Structure Science, KEK. 
\bibliography{LVO1tex}

\end{document}